# Unidirectional Oriented Water Wire in Short Nanotube

Le Jin[a], Depeng Zhang[a,b], Yu Zhu[a], Xinrui Yang[a,c], Yi Gao[d], Zhigang Wang[a,1]

[a]Institute of Atomic and Molecular Physics, Jilin University, Changchun 130012, China;

[b]Institute of Molecular Sciences and Engineering, Institute of Frontier and Interdisciplinary Science, Shandong University, Qingdao 266237, China;

[c]College of Physics, Jilin University, Changchun 130012, China;

[d]Shanghai Advanced Research Institute, Chinese Academy of Sciences, Shanghai 201210, China;

[1]Correspondence authors: wangzg@jlu.edu.cn (Z. W.).

Abstract: The orientation of water molecules is the key factor for the fast transport of water in small nanotubes. It has been accepted that the bidirectional water "burst" in short nanotubes can be transformed into unidirectional transport when the orientation of water molecules is maintained in long nanotubes under the external field. In this work, based on molecular dynamics simulations and first-principles calculations, we showed without external field, it only needs 21 water molecules to maintain the unidirectional single file water intrinsically in carbon nanotube at seconds. Detailed analysis indicates that the surprising result comes from the step by step process for the flip of water chain, which is different with the perceived concerted mechanism. Considering the thickness of cell membrane (normally 5-10 nm) is larger than the length threshold of the unidirectional water wire, this study suggests it may not need the external field to maintain the unidirectional flow in the water channel at the macroscopic timescale.

## Introduction

The orientations of confined water molecules are crucial for the fast transport of water in channels (1-3). Since the discovery of water flux in thin carbon nanotubes in 2001, the extensive studies have been devoted to the extraordinary water behaviors in confined space for their potential applications in the artificial water channel and desalination (1, 4-10). Using the carbon nanotubes as the distinct model systems, many factors have been identified to influence the water transport in small channels, such as the defect of water wires, the diameter of nanotubes, and etc (11-13). At present, it is well accepted that without the external field, there are bidirectional water flux in the short nanotubes because of the fast and concerted flip of the orientation of the water chain (1). To realize the unidirectional flow for practical applications, a number of intricate strategies have been proposed, such as the induced external charge or force (14-16). Despite the above significant progress, the fundamental understanding of the flip of hydrogen bonds of confined water is still very limited due to the lack of the studies at the atomic level.

In this work, combining the classical simulations and first-principles calculations, we demonstrated the flip of hydrogen bonds for the single file water chain within carbon nanotube is a step by step process. The persistent time of dipole alignment for water chain is exponential to the number of the water molecules. In particular, it takes only 21 water molecules to maintain the orientation of water chain at seconds. As the unidirectional orientation of water chain is the prerequisite of the unidirectional water flow, this study suggests length of channel for the unidirectional water flow is much shorter as we perceived before. More important, considering the thickness of cell membrane (normally 5-10 nm) is larger than the length threshold of the unidirectional water wire, there may suggest the spontaneous water flow in the water channel at the macroscopic timescale.

## Result and discussion

To study the flip process of water chain under the confinement, the molecular dynamics (MD) simulations were first performed using 1-7 $H_2O$ molecules in (6, 6) CNT. Here, we used 2, 3, 4 $H_2O$ molecules as the example, and other cases were given in the Supplementary Materials. The water molecules in the CNT form the one-dimensional chain (Fig. 1A). And the angle between the dipole and plane perpendicular to the axis of the CNT is used to determine the orientation of the water molecules (Fig. 1B). Fig. 1C gives the evolution of dipole strength with the simulation time for the water chain. For all the systems, the dipole orientations flip frequently. The average time between the flip intervals of water chain are about 0.8 picosecond (ps) for dimer, 6.0 ps for trimer, and 21 ps for tetramer, respectively. It is clear that the increase of the number of water molecules slower the dipole flips of the water chain, and the duration time $t$ for each direction becomes longer significantly. To describe the lifetime of the direction of the axial dipole, the average duration time $t$ can be expressed using an exponential equation fitted as $Log(t)=a*N+b$, as shown in Fig. 1D. According to the relationship, when the number of water molecules was extrapolated to 12, 16 and 21, the duration time required to maintain the direction of axial dipole could reach microsecond (μs), millisecond (ms) and second (s), respectively.

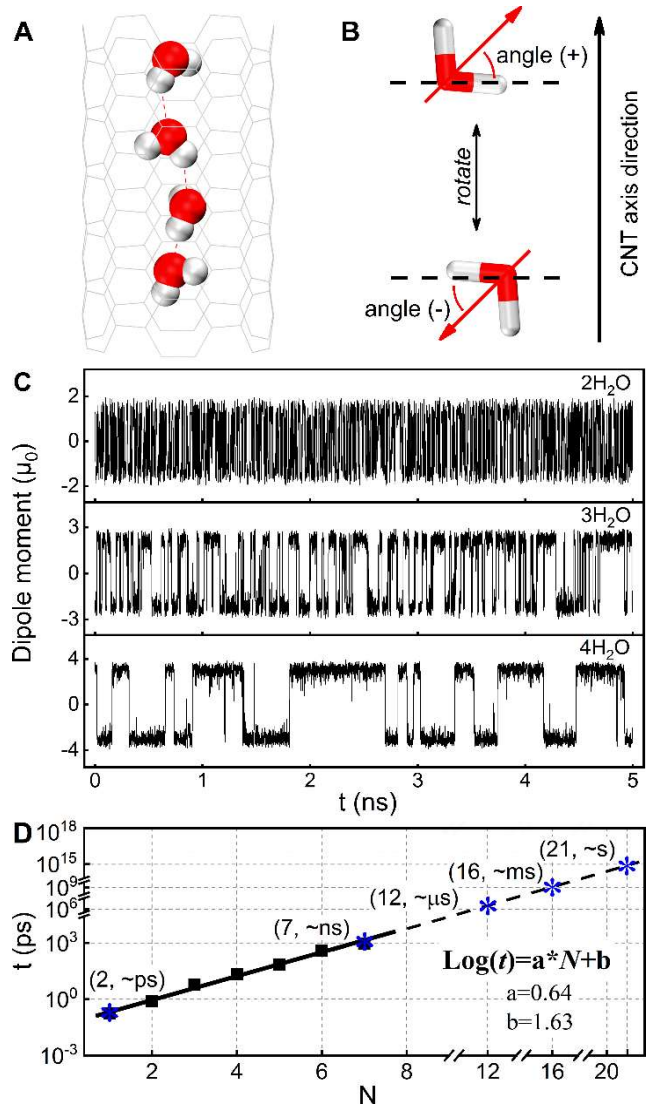

**Fig. 1.** Effect of the number of water molecules on persistent time of dipole alignment for water chains. (A) The model of confined water chain. (B) The definition of molecular dipole angle in water chains. (C) The change of the dipole moment with the time. Here, the water chain is consist of 2, 3, 4 water molecules, respectively. The $\mu_0$ represents the dipole moment of one water molecule. (D) The relationship between the lifetime of the orientation of the axial dipole moments and the number of water molecules. The coefficient a and the constant b are 0.64 and 1.63 for our simulations, respectively. The N is the number of water molecules. Solid square points are the data obtained by MD simulations. The line is obtained by fitting the former. It can be express as the function shown in the figure. The dashed part of the line represents the extrapolation based on the fitted relationship. The $10^3$, $10^6$, $10^9$, $10^{12}$, and $10^{15}$ represent picosecond (ps), nanosecond (ns), microsecond (μs), millisecond (ms) and second (s), respectively. The "~" means magnitude.

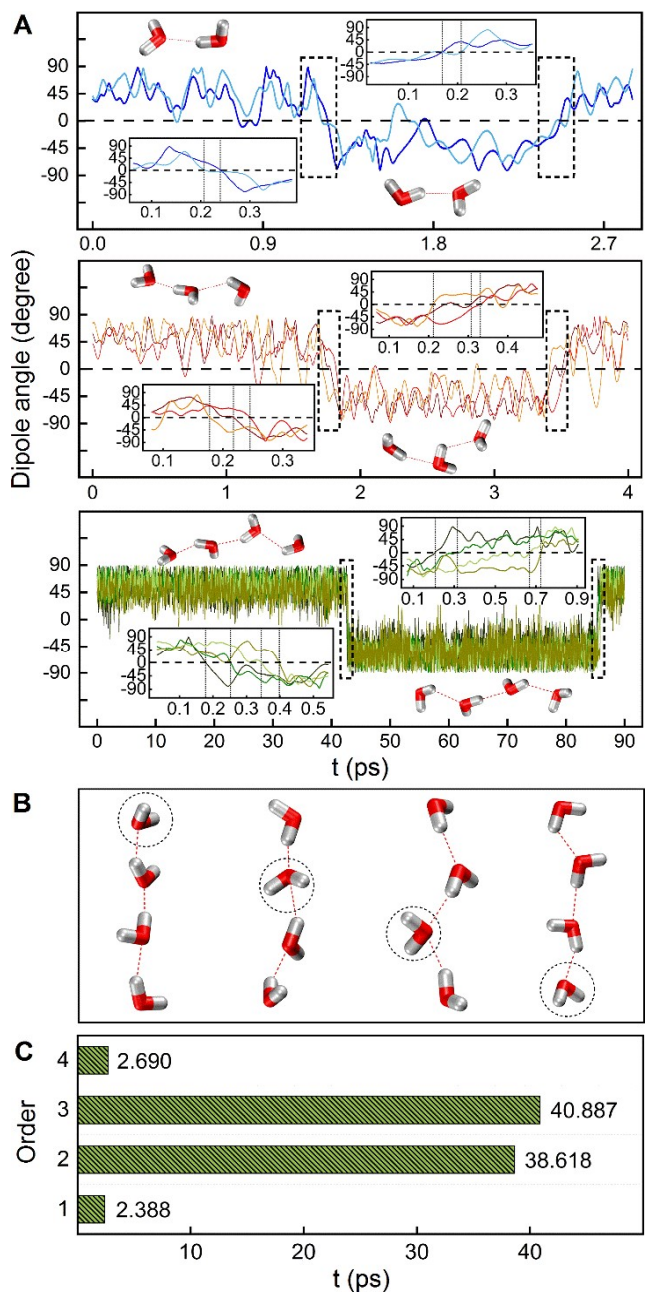

**Fig. 2.** The snapshots of trajectories for the typical molecular orientation of water chains. (A) The change of dipole angle for each water molecule in the different chain. Here, different color indicates different number of water molecules. (B) The flip process of water tetramer chain. (C) Comparison of the rotation time of each water molecule for the tetramer chain.

Further, to understand the underlying mechanism of the flip of the water chain in the nanotube, the snapshots of trajectories for the typical orientation of water molecules are given in Fig. 2. In Fig. 2A, it is clear that the angles fluctuate within the duration time. When the number of $H_2O$ molecules increases, the fluctuation time of the molecules become longer. And the fluctuations normally correspond to the flips of the molecules, which do not trigger easily the flip of the water chain. We can also find that the flip of water molecules does not occur simultaneously (inset of Fig. 2A). The flip of the water chain is a step by step process rather than a concerted process as perceived in previous literatures. The flip process of tetramer chain in Fig. 2B clearly shows that the one-by-one flip of the water molecule from the one end to the other end. Moreover, as shown in Fig. 2C, the average rotation time for end $H_2O$ molecules takes only few picoseconds, but that for the middle $H_2O$ molecule takes up to tens of picoseconds. This shows the both ends of water molecules of the chain rotate more easily than that of the internal water molecules. In order to confirm the process more, the snapshots of the trajectories for water pentamer chain were also analyzed, which is also consistent with the feature of water tetramer chain. (See Fig. S3 for details).

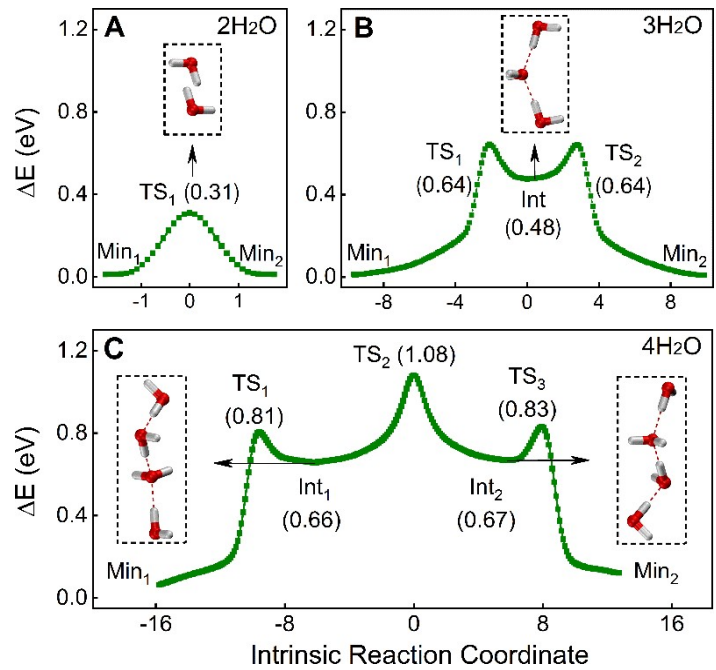

**Fig. 3.** The potential energy surfaces (PESs) of flip processes for water chains. The water chain structures in the dotted frame corresponds to the transition states (TSs) or intermediates. The water chain is consist of (A) dimer, (B) trimer and (C) tetramer.

To reveal the physical mechanism of flip of the water chain, the first-principle calculations were performed. Using 2-4 $H_2O$ molecules in (6, 6) CNT as the model systems, the transition states (TS) and intrinsic reaction coordinates (IRCs) were analyzed. As shown in Fig. 3, the flip process of the water chain generates a series of defective structures (11), which can be considered as the intermediates of reaction path. The formation of intermediates is an essential step in the process of water chain flip. This is significantly different from what was previously considered defective structures (8, 17).

Remarkably, there is only one TS and no intermediate for the water dimer chain. For the water trimer chain, there are two TSs and one intermediate. When the number of water molecules is four, the numbers of TSs and intermediates becomes 3 and 2, respectively. The number of TSs and intermediates increases with the increase of the number of water molecules. It is reasonable to infer that, the number of TSs and intermediates would be *N-1* and *N-2*, if the number of water molecules is *N*. That is to say, when the number of water molecules exceeds 2, the flip process not only be accompanied by the dipole flip, the breaking and generating of H-bonds, and the conversion of donor and acceptor roles between molecules (18), but also contain the formation and disappearance of intermediates.

Meanwhile, it should be noted that the energy barrier to be overcome is closely related to the number of water molecules. Just as shown in Fig. 3, for the water trimer, the energy barrier is about 0.64 eV. It is about 0.33 eV higher than that of water dimer chain. For the tetramer water chain, it is at least 0.5 eV higher than that of two water molecules. It is clear that the elongation of the water chain restrains the rotation of the water molecules, especially for those in the middle of the water chain. This can be attributed to the enhanced dipole moments with the increased number of water molecules (Fig. 1). In addition, we can also notice that the rotation of the end molecule also tends to flipping back to the initial structure rather than to simply induce the rotation of the next $H_2O$ molecule (using $(H_2O)_4$ as the example), because the Int1 has the much lower backward barrier (0.15 eV) to TS compared to the forward barrier (0.42 eV) to Int2. This further demonstrates the flip of the water chain is a step by step process, which means that the difficulty of water chain flip is mainly caused by the interaction between water molecules. This provides favorable support for water chain self-drive.

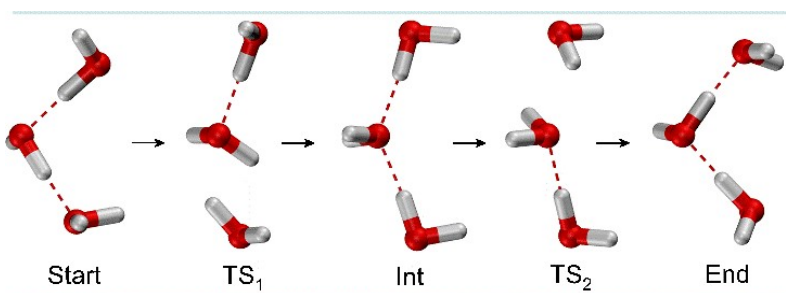

**Fig. 4.** The picture of flip process for the water trimer chain in CNT.

The detailed flip process of water trimer is given as an example in Fig. 4. It shows the change of the H-bonds orientations when the water chain flips. The five structures correspond to the five extreme points on the PES. In detail, the initial orientations of all H-bonds are upward. Then, from reactant to intermediate, the H-bond of the lower two water molecules is broken. When the H-bond is rebuilt, its orientation becomes downward. However, there is no change for the H-bond orientation of the upper two water molecules. While from intermediate to product, the H-bond between the upper two water molecules is broken and rebuilt, and the orientation becomes downward. For the lower two water molecules, the H-bond orientation is still downward. Therefore, the complete flip process of the water chain is demonstrated. For the flip details of two water molecules and four water molecules, see Fig. S (4-5). The analysis above all shows that the defective structures are the stable intermediates, which retards the stepwise flip of the water chain in CNT and prefer retaining the original orientation of the dipoles.

## Conclusion

In this paper, our work suggests that the single-file water chain could retain its original orientation up to seconds at the extremely short CNT at less than 5 nm. The defective structures in the water chain can be considered as the stable intermediates during the flip process, which are the necessary pathways for water chain flip rather than accidentally generated. Further, a new flip picture is established. The stepwise flip of the water molecules in the single file water chain is contrary to the previous presumption of the concerted mechanism, which induces the unexpected maintenance of the unidirectional orientation of water chain. Considering the unidirectional orientation of the water molecules is the prerequisite of the unidirectional water flow and the water channels of the cell membrane are normally longer than 5 nm, our discovery indicates the transmembrane water channel might maintain the unidirectional water flow at the macroscopic timescale without the external field.

## Methods

The molecular dynamics simulations were performed at the temperature of 300 K in Gromacs 4.6.7 program (19). The ensemble and thermostat were set as NVT and Berendsen, respectively. The confined system was under the periodic boundary conditions with a box size of 25 Å × 25 Å × 39.352 Å. TIP3P water model was employed to form the water chain in the CNT. The carbon atoms of CNT were modeled as uncharged Lennard-Jones particles to ensure that only the van der Waals effects on the water. A time-step of 1 fs was used, and data were also collected every 1 fs. The time for the simulation was 105 nanoseconds (ns) for each process and the last 100 ns of simulation were collected for analysis.

For first-principles calculations, the empirical-dispersion-corrected hybrid Perdew-Burke-Ernzerhof (PBE0-D3) method of density functional theory (DFT) was carried out in the Gaussian 09 package (20). The Basis sets 6-311+G(d, p) and 6-31G(d) were used for water and CNT, respectively. The diameter, length of the armchair type single-walled (6, 6) CNT simulated by us are 8.20 Å, 20 Å. In calculations except for the CNT pre-optimization, all atoms of the CNT were frozen to provide the constant confinement effect. The water chain was confined in the CNT along the tube axis, and these water molecules were connected with each other by H-bonds. From different initial geometries of confined water chains, we searched the structures of extreme points (including equilibrium and transition state) in the rotational process of water molecules in the CNT, and traced reaction paths of flip process of water chains by intrinsic reaction coordinate (IRC).

## ACKNOWLEDGMENTS
This work was supported by the National Natural Science Foundation of China (under grant number 11974136 and

11674123). Z. W. also acknowledges the High-Performance Computing Center of Jilin University and National Supercomputing Center in Shanghai.